\documentclass[aps,prd,twocolumn, superscriptaddress,nofootinbib]{revtex4}
\usepackage{fancyhdr}
\usepackage{wrapfig}
\usepackage{amsmath,amsthm,amssymb}
\usepackage{enumerate}
\usepackage{color}
\usepackage{graphicx}
\usepackage{caption}
\usepackage{subcaption}
\usepackage{sidecap}

\def\beq{\begin{equation}}
\def\eeq{\end{equation}}
\def\bea{\begin{eqnarray}}
\def\eea{\end{eqnarray}}

\def\bwt{\begin{widetext}}
\def\ewt{\end{widetext}}

\newcommand*{\dd}{\mathrm{d}}

\begin{document}

\title{Cosmic Censorship in Lorentz Violating Theories of Gravity}

\author{Michael Meiers}\email{mmeiers@uwaterloo.ca}
\affiliation{Department of Physics and Astronomy, University of Waterloo, Waterloo, ON, N2L 3G1, Canada}
\author{Mehdi Saravani}\email{msaravani@pitp.ca}
\affiliation{Perimeter Institute
for Theoretical Physics, 31 Caroline St. N., Waterloo, ON, N2L 2Y5, Canada}
\affiliation{Department of Physics and Astronomy, University of Waterloo, Waterloo, ON, N2L 3G1, Canada}
\author{Niayesh Afshordi}\email{nafshordi@pitp.ca}
\affiliation{Perimeter Institute
for Theoretical Physics, 31 Caroline St. N., Waterloo, ON, N2L 2Y5, Canada}
\affiliation{Department of Physics and Astronomy, University of Waterloo, Waterloo, ON, N2L 3G1, Canada}

\begin{abstract}
Is  {\it Cosmic Censorship} special to General Relativity, or can it survive a violation of equivalence principle?
Recent studies have shown that singularities in Lorentz violating Einstein-Aether (or Horava-Lifhsitz) theories can lie behind a {\it universal horizon} in simple black hole spacetimes. Even infinitely fast signals cannot escape these universal horizons. We extend this result, for an incompressible aether, to 3+1d dynamical or spinning spacetimes which possess inner killing horizons, and show that a universal horizon always forms in between the outer and (would-be) inner horizons. This finding suggests a notion of {\it Cosmic Censorship}, given that geometry in these theories never evolves beyond the universal horizon (avoiding potentially singular inner killing horizons).   
A surprising result is that there are 3 distinct possible stationary universal horizons for a spinning black hole, only one of which matches the dynamical spherical solution. This motivates dynamical studies of collapse in Einstein-Aether theories beyond spherical symmetry, which may reveal instabilities around the spherical solution. 

\end{abstract}
\maketitle

\section{Introduction}

Theory of general relativity (GR) has been successful in describing a wide range of phenomena, from solar system to cosmological scales. In addition to being consistent with various experiments, the mathematical elegance of the theory is very appealing. Diffeomorphism invariance, at the core of GR, gives a straightforward constructive way of building the theory. In fact, GR is the simplest diffeomorphism invariant theory for metric.

From observational point of view, there is no reason to abandon this theory. GR is compatible with a wide variety of experimental constraints\footnote{Although there have been various attempts to solve the problems of dark matter and dark energy with GR modifications, simple solutions to these problem in the context of GR exist. In other words, there is no apparent observational contradiction with GR which necessitates GR modifications.}. On the other hand, many attempts have shown so far that modifying GR is a tricky task, and one often faces physically unacceptable results, e.g. the appearance of Boulware-Deser ghost in massive gravity \cite{Boulware:1973my} and ghost degrees of freedom in quadratic gravity \cite{Stelle:1976gc}. 

However, studying non-GR theories of gravity is still valuable, and  the main reason stems from quantizing gravity. GR, while being a very successful classical theory, has failed to cope with quantum mechanics. 
Therefore, one approach to quantum gravity has been to abandon diffeomorphism invariance (at high enough energies), as e.g., done in the celebrated Horava-Liftshitz gravity \cite{Horava:2009uw}.

In different examples of theories with broken Lorentz invariance, superluminal degrees of freedom appear (see \cite{Blas:2010hb,Jacobson:2000xp}). The existence of superluminal excitations (SLE) points out that a different causal structure exists in these theories compared to GR, even when the back-reaction of these excitations on the geometry is negligible. This property is especially of significance in the black hole (BH) solutions. While potentially SLE can escape the traditional killing horizon of a BH and make the classical theory unpredictable, it has been shown in many examples \cite{Sotiriou:2014gna,Ding:2015kba,Barausse:2012qh,Bhattacharyya:2015gwa,Barausse:2013nwa,Babichev:2006vx,Barausse:2011pu,Blas:2011ni} that a notion of horizon (called universal horizon) still exists in these theories. Moreover, universal horizon (UH) thermally radiate and satisfies the first law of horizon thermodynamics\footnote{so far only for spherically symmetric solutions} \cite{Mohd:2013zca,Berglund:2012bu,Berglund:2012fk,Bhattacharyya:2014kta}. Studying the notion of universal horizon and its temperature and entropy is important since it guides us to better understanding the structure of UV theory.

In this paper, we study the universal horizon formation in dynamical or stationary spacetimes with an inner killing horizon, in the limit of infinite sound speed for excitations (i.e. {\it incompressible} limit). In order to do so, we make use of the fact that surfaces of global time (defined by the background field), in the incompressible limit, coincide with constant mean curvature (CMC) surfaces of the spacetime. Furthermore, the backreaction of the incompressible field on the spacetime geometry is negligible as long as the event horizon is much smaller than the cosmological horizon  \cite{Saravani:2013}. In the next section, we show how the universal horizon forms in a dynamical setting, in the collapse of a charged shell, and we derive a formula for the radius of the universal horizon in terms of the charge. In Section \ref{geoUH}, we propose a geometric definition for universal horizon. This allows us to study the universal horizon for spinning black holes. In \ref{KerrUH} we show that there are three axisymmetric surfaces which satisfy the conditions of a universal horizon. As we show, this means that two families (with infinite numbers) of axi-symmetric universal horizons in Schwarzschild case exist.
Section \ref{conclude} concludes the paper.

\section{Universal Horizon in Dynamic Reissner--Nordstrom Geometry}
\label{RN}
We start this section by finding CMC slicing of dynamic Reissner--Nordstrom (RN) geometry. As we mentioned earlier, CMC surfaces of this spacetime are the constant global time surfaces of the background incompressible field, and they define the new causal structure imposed by this field (see the analysis in \cite{Saravani:2013}). Once we derive the CMC slicing, we focus on the (universal) horizon formation in this geometry.

\subsection{CMC Surfaces in a Dynamic Reissner--Nordstrom Geometry}
\label{CMC-RN}

In order to examine the formation of the universal horizon in a dynamic Reissner--Nordstrom geometry, one must first describe surfaces of constant mean curvature for a collapsing charged massive spherical shell. An examination of CMC surfaces has been similarly looked at in the restricted case of maximal surfaces ($K=0$) \cite{Maeda:1980}. The dynamics of the collapse itself is well known and described by Israel \cite{Israel:1967}. Describing the metric in the standard way has the geometry inside the shell as flat and the RN outside. We write this geometry as:

\begin{align*}
  &\dd s^2 =f_-(r) \dd t_-^2 -f_-(r)^{-1} \dd r^2 -r^2\dd \Omega^2  &(r<R)\\
  &\dd s^2 = f_+(r)\dd t_+^2 - f_+(r)^{-1}\dd r^2 -r^2\dd \Omega^2 &(r>R)
\end{align*}
where $f_-(r)=1$ and  $f_+(r)=1-\frac{2M}{r}+\frac{Q^2}{r^2}$ in $G=c=1$ units. The parameters are the gravitational mass $M$ and shell's charge $Q$. For simplicity we will often use relative charge $\mathit{q}=Q/M$. While the spherical coordinates are shared between the inner and outer regions, the time coordinates $t_-$ and $t_+$ correspond to the Minkowski and RN time respectively.

Let the family of spacelike CMC surfaces be denoted by $\Sigma_K(t_g)$ where $t_g$ is a global time coordinate that is constant for each surface. The timelike normal vector to this surface is labelled $v^\mu$. The CMC condition implies $\nabla_\mu v^\mu=K$, resulting in:
\beq\label{CMC_eq}
\frac{\partial}{\partial t_\pm}v^{t_\pm}+\frac{1}{r^2}\frac{\partial}{\partial r}r^2v^r =K
\eeq

If we denote $B \equiv -r^2v^r$ and use the normalization condition $v_\mu v^\mu =1$ then:
\beq
r^2v^{t_\pm}=\pm f_\pm(r)^{-1}\sqrt{B^2+f_\pm r^4}.
\eeq
For now we use the '$+$' case so that for $v_t>0$ for $r\gg M$. Additional explanation and the cases where '$-$' is relevant will be seen in Section \ref{foliation_struc}.  Combining this result with \eqref{CMC_eq} we get

\begin{equation}
  \label{eq:pde}
   \frac{B}{f_\pm(r)\sqrt{h(r,B)}}\frac{\partial}{\partial t_\pm}B-\frac{\partial}{\partial r}B=Kr^2
\end{equation}

with $h(r,B)=B^2+f_\pm r^4$. The characteristic equations of (\ref{eq:pde}) are simply:

\begin{equation}
  \label{eq:chareq}
   \frac{\dd t_\pm}{\dd s}= \frac{B}{f_\pm(r)\sqrt{h(r,B)}},~\: \frac{\dd r}{\dd s}=-1, \: ~\text{and}~ \: \frac{\dd B}{\dd s}= K r^2,
\end{equation}
for some parameter $s$. Using the second equation of (\ref{eq:chareq}) to integrate the first and third equations results in:

\begin{equation}
   \label{eq:coordeq}
   t_\pm=(t_\pm)_0-\int_{r_0}^r \frac{B\dd r}{f_\pm(r)\sqrt{h(r,B)}},~~ \text{and}~~ B= \frac{K}{3} (r^3-r_0^3)+B_0,
\end{equation}
 where $(t_\pm)_0$, $r_0$, and $B_0$ are integration constants. In order to fix these constants, we examine the internal and external cases separately. 

\paragraph{1) Inside the shell:}

If $r_0=0$ then $B(r=0)=B_0$. If $B_0 \neq 0$ this would lead to a contradiction, as $v^r=\frac{-B}{r^2}$ should be finite in the flat geometry. Therefore with $r_0=0$, the equation reduces to:

\begin{equation}
  \label{eq:coordeqinside}
   t_-=(t_-)_0+\int_{0}^r \frac{Kr^3\dd r}{3\sqrt{(\frac{Kr^3}{3})^2+r^4}} \: \text{and}\: B= \frac{K}{3} r^3
\end{equation}

\paragraph{2) Ouside the shell:}

Let $r_0=R((t_+)_0)$, we can determine $B_0$ by looking at the boundary between the flat and RN spaces. Projecting the vector $v^\mu$ along the shell should give us continuous observable values. The shell timelike path comes from $S=R(t_\pm)-r=0$ which creates the unit normal vector and tangent vector labelled as $n^\mu$ and $u^\mu$ respectively. If we choose the sign of the normalization factors such that $u^r<0$ , the vectors take the form of:
\bea
 n_-^\mu&&=\frac{g^{\mu\nu}}{N_-}(\nabla_-)_\nu S=\frac{1}{N_-}(\frac{\dd R}{\dd t_-},1,0,0),\\ 
u_-^\mu &&=\frac{1}{N_-}(1,\frac{\dd R}{\dd t_-},0,0),\\
N_-^2&&=1-(\frac{\dd R}{\dd t_-})^2,
\eea
inside the shell, while outside takes the from of:
\bea
n_+^\mu&&=\frac{g^{\mu\nu}}{N_+}(\nabla_+)_\nu S=\frac{1}{N_+}(f_+^{-1}\frac{\dd R}{\dd t_+},f_+,0,0),\\
u_+^\mu&&=\frac{1}{N_+}(1,\frac{\dd R}{\dd t_+},0,0),\\
N_+^2&&=\frac{f_+^2-(\frac{\dd R}{\dd t_+})^2}{f_+}.
\eea
We wish to find functions $C(R)$ and $D(R)$, such that:
\beq
v_-^\mu=C n_-^\mu+D u_-^\mu.
\eeq
 From inside the shell $v^\mu(R)=(1,0,0,0)$ which means $C=\frac{-1}{N_-}\frac{\dd R}{\dd t_-}$ and $D=\frac{1}{N_-}$. Requiring projections ($C$ and $D$) to be the same from outside, we get: 
\beq
B_0=-R^2(C (n_+)^r+D (u_+)^r)=\frac{R^2}{N_+N_-}\left(\frac{\dd R}{\dd t_+}-f_+ \frac{\dd R}{\dd t_-}\right).
\eeq
So, if we specify the dynamics of the shell $\frac{dR}{dt_{\pm}}$, all the parameters are fixed.

The description of the radial velocity comes from Israel and De La Cruz \cite{Israel:1967}:
\bea
\left( \frac{\dd R}{\dd t_-}\right)^2&&=
1-\frac{R^2}{(\epsilon R-b)^2}\label{drdt+},\\
\left( \frac{\dd R}{\dd t_+}\right)^2&&=
f_+^2-\frac{f_+^3 R^2}{(\epsilon R-b-\frac{m}{\epsilon})^2},\label{drdt-}
\eea
where $\epsilon=\frac{M}{\mathcal{M}}$ and $b=\frac{M(\epsilon^2\mathit{q}^2-1)}{2\epsilon}$ with $\mathcal{M}$ denoting the total rest mass. 
We can use (\ref{drdt+}) and (\ref{drdt-}) to reduce $N_+=\frac{Rf_+}{\epsilon R-b-\frac{M}{\epsilon}}$ and $N_-=\frac{R}{\epsilon R-b}$. Note that $N_+$ changes signs to enforce $u^r<0$, becoming negative only when $\frac{\dd R}{\dd t_\pm}$ flips signs. These choices  simplifies $B_0$ to:
\beq
B_0=\frac{M}{\epsilon}\sqrt{(\epsilon R-b)^2-R^2}.
\eeq

\begin{figure}
  \centering
    \includegraphics[width=0.5\textwidth]{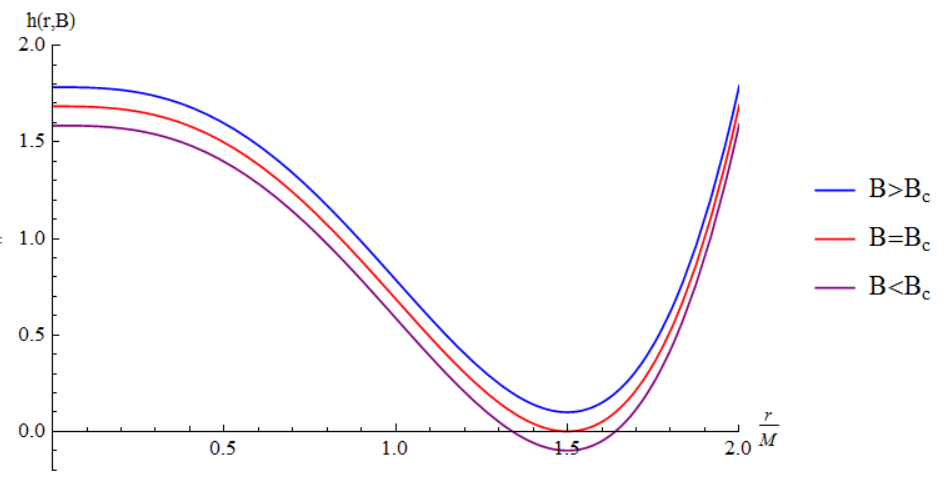}
  \caption{$h(r,B)$ with sub-critical, post-critical and critical $B$.}
  \label{fig:hsols}
\end{figure}

\subsection{Horizon Formation}

Following the analysis of \cite{Saravani:2013}, we examine the properties of $t_+$. The behaviour of $t_+$ heavily depends on $h(r,B)$. While $B$ is large, which corresponds to large $R$, $h(r,B)$ is never vanishing. However when a critical value $B_c$ is reached, $ h(r,B_c)$ has double root at a particular value of $r$ labelled $r_h$ (see Figure \ref{fig:hsols}). Something interesting will occur when $r_h$ is larger than the radius of the shell for which $B_c$ occurs named $R_{lc}$ or radius of {\it last contact} ($B(R_{lc})=B_c$). A signal sent out from the shell at $R_{lc}$ will proceed out to $r_h$ but  takes infinitely long time to ever reach this radius. In fact, signals sent just outside $R_{lc}$ will form an envelope around $r_h$ staying at this radius longer and longer, as $R_{lc}$ is approached, before escaping to infinity (see Figure \ref{fig:plot0log}). The values of $B_c$ and $r_h$ can be found by finding the solutions to $h(r,B)=\frac{\partial h(r,B)}{\partial \mathrm{r}}=0$. 
We examine this equation in two different cases. 

\subsubsection*{Case 1: $K=0$}
 
Equations for the double root reduce to: 

\bea
&&r_h^4-2Mr_h^3+Q^2r_h^2+B_c^2=0\label{eq1_rh}\\
&&2r_h^3-3Mr_h^2+ Q^2r_h=0\label{eq2_rh}
\eea

to which the solutions with non-negative real $B_c$ are the trivial $r_h=B_c=0$ and
\bea
r_h&&=\frac{3M}{4}+\frac{M}{4}\sqrt{9-8\mathit{q}^2} \label{r_h_ns}\\
B_c&&=r_h\sqrt{-r_h^2+2mr_h-Q^2}=r_h^2\sqrt{-f_+(r_h)}.
\eea

\begin{figure}[b]
  \centering
    \includegraphics[width=0.5\textwidth]{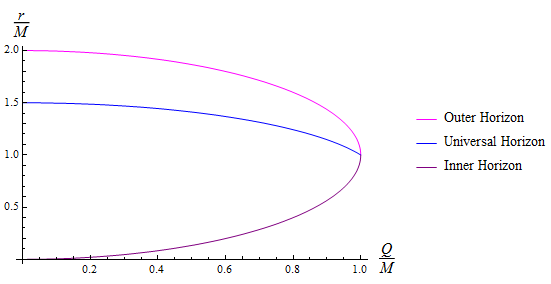}
  \caption{The outer, inner, and universal horizons for $K=0$ and varying $Q$ }
  \label{fig:uni-hor}
\end{figure}

It is of interest to not that the UH is always between the inner and outer killing horizons of the metric (see Figure \ref{fig:uni-hor}).

\subsubsection*{Case 2: $K\not=0$ }

Equations for the double root are written as:
\bea
&&\frac{K^2}{9}r_h^6+r_h^4+(\frac{2 K B_0}{3}-2M)r_h^3+Q^2r_h^2+B_0^2=0,\notag\\
&&\frac{ K^2}{3}r_h^5+ 2r_h^3+(k B_0-3M)r_h^2+ Q^2r_h=0.\notag
\eea

\begin{figure}[b]
  \centering
    \includegraphics[width=0.5\textwidth]{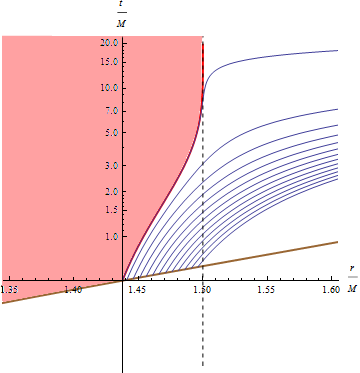}
  \caption{The universal horizon formation for $Q=0$ in Schwarzschild coordinates. The blue lines represent CMC surfaces, the lowest brown line is the shell's surface, the red line is the universal horizon (UH) and the dotted black is the radius that UH asymptotes to. Here, and in all the subsequent diagrams, the red region lies behind the UH.  }
  \label{fig:plot0log}
\end{figure}

The non-trivial solution for $B_c$ is:
\beq
B_c=r_h\sqrt{-r_h^2+2Mr_h-Q^2}-\frac{Kr_h^3}{3}=r_h^2\sqrt{-f_+(r_h)}-\frac{Kr_h^3}{3},
\eeq
however the solution for $r_h$ can be at best expressed perturbatively in $K$. To linear order the solution is:

\beq
r_h=r_h^0-K\frac{(r_h^0)^3\sqrt{-f_+(r_h^0)}}{M\sqrt{9-8q^2}}+\mathcal{O}(K^2),
\eeq
where $r_h^0=\frac{3M}{4}+\frac{M}{4}\sqrt{9-8\mathit{q}^2}$. Assuming that the expansion of the background field is negligible (for example fixed by cosmology, as $K=3 \times$Hubble constant), in the region of interest $0<r<2M$ the effect of terms containing $K$ are insignificant. From here we set $K=0$ as its effect will only come into play when looking at the causal structure when $R$ is very large. 

The last unknown of the universal horizon is where it begins before it asymptotes to $r_h$. To solve for $R_{lc}$  we use that $B(R_{lc})=B_c$ and it results in either: 

\beq
R_{lc}=\frac{b}{2}-\frac{B_c^2}{2 b M^2},
\eeq
for $\epsilon=1$ or
\beq
R_{lc}=\frac{\epsilon b+\sqrt{\epsilon^2(\epsilon^2-1)B_c^2/M^2+b^2}}{\epsilon^2-1},
\eeq
for $\epsilon >1$.

\begin{figure}
\begin{subfigure}[t]{0.45\textwidth}
\centering
\includegraphics[width=\hsize]{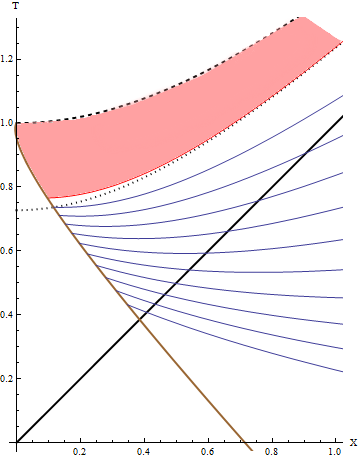}
\caption{The universal horizon formation for $Q=0$ in Kruskal-Szekeres coordinates. The coloured lines/region have the same meaning as Figure \ref{fig:plot0log}, additionally the thick black line represents the null horizons and the dashed line the singularity.  }
\label{fig:plot0kr}
\end{subfigure}
\begin{subfigure}[t]{0.5\textwidth}
\centering
    \includegraphics[width=\hsize]{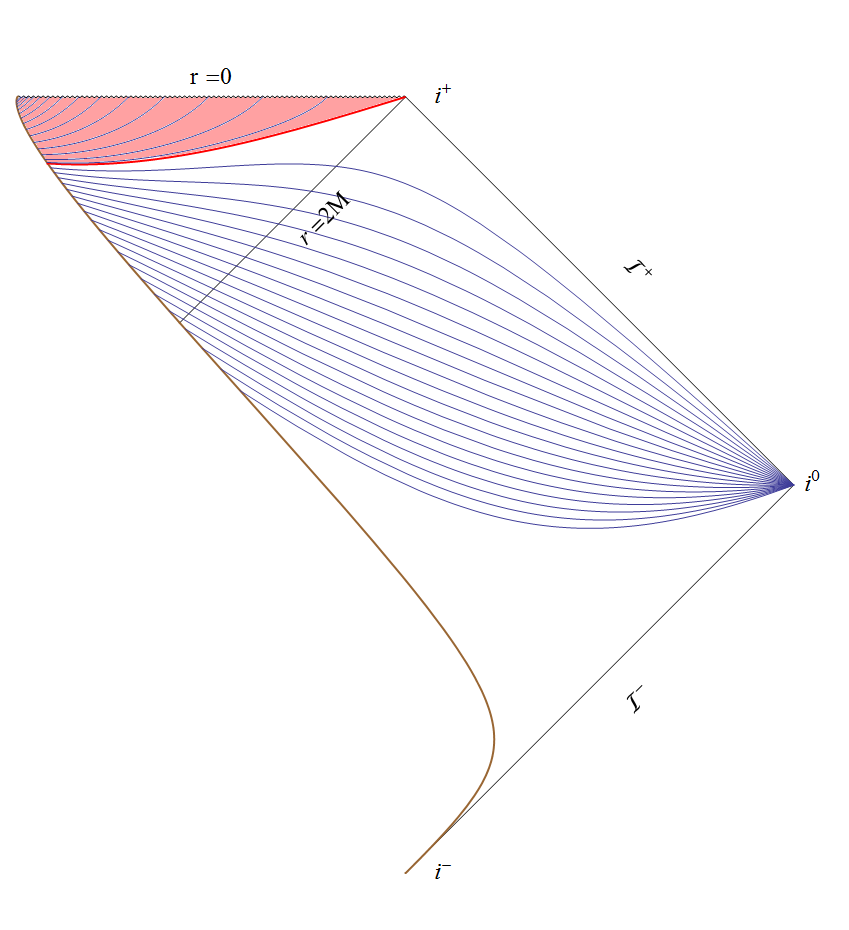}
  \caption{The Penrose diagram for a $Q=0$ collapsing shell depicting the UH horizon. Coloured lines/region have the same meaning as Figure \ref{fig:plot0log}  with the inclusion of sub-UH CMC surfaces in blue.
}
\label{fig:plot0pn}
\end{subfigure}
\caption{Surfaces for constant global time and formation of the universal horizon in Kruskal-Szekeres coordinates and Penrose diagram for $Q=0$.}
\end{figure}



We take the larger of the solutions to the quadratic, as the first instance of $B_c$ will create the behaviour desired.

\subsection{Inside the Universal Horizon}

The foliation can be extended for $B<B_c$, however some subtleties arise. Denote the unit tangent vector of the CMC surfaces at the point the surface intersects the shell as $s^\mu$ which in components can be written as:
\bea
s^\mu&&=\frac{1}{N_s}(T_{cmc}'(R),1,0,0),\\
N_s^2&&=\frac{1-f_+^2T_{cmc}'^2}{f_+}=\frac{R^4}{h(R,B(R))},
\eea
where the last equality comes from using the derivative of \eqref{eq:coordeq}. In general $h(r,B(R))$ is a quartic that can not easily be factored, however when restricted to the surface of the shell it can be factored to:
\beq
h(R,B(R))=\left(R^2-MR+\frac{Mb}{\epsilon}\right)^2.
\eeq

Thus one can write the normalization factor as: 
\beq
N_s=\frac{R^2}{R^2-MR+\frac{Mb}{\epsilon}}.
\eeq
As a result, between the zeroes of $1/N_s$ at $M\frac{1\pm\sqrt{1-4b/\epsilon}}{2}$ we get $s^r<0$. In particular this means that rather than increasing in $r$ the CMC surfaces that intersect between these two roots will have a strictly decreasing $r$ coordinate. Moreover it is precisely at these points where $s^t$ switches signs corresponding to the second solution for $T'_{cmc}$ which comes from the '-' solution to $r^2v^{t_+}$. 

The second complication occurs when $R$ does not lie between the roots of $N_s$ while still being less than $R_{lc}$. Here the CMC surface increases its radial coordinate initially only to encounter a zero of $h(r,B(R))$ at $r_{turn}$. The integral for $t_+$ can carried out since $T'_{cmc}$ only depends on the square root of $h(r,B(R))$ and, by construction, $r_{turn}$ is only a first order zero of $h(r,B(R))$. After this point $T'_{cmc}$ switches signs and the $r$ coordinate begins decreasing, flipping the direction of the integration taking the surface from $r_{turn}$ to $0$. 
 
Now that these subtleties are understood, we are ready to examine the structure of the complete foliation.

\subsection{Foliation Structure}\label{foliation_struc}
We will break up the discussion into several sections. For all our analysis  we consider the shell to be dropped from infinity thus $\epsilon\geq1$ \cite{Israel:1967}, in particular the inward velocity of the shell at infinity is exactly $\sqrt{\epsilon^2-1}$. There are  4 cases of interest:%

\subsubsection{Case 1: $Q=0$}
When restricted to Schwarzschild, $b=\frac{-M}{2\epsilon}$  making it strictly negative. In particular the value of $\epsilon$ is only relevant to the radius of last contact, and so without losing any depth of examination we set $\epsilon=1$. Figures \ref{fig:plot0log} and \ref{fig:plot0kr} illustrate the foliations created by the CMC surfaces for $R>R_{lc}$ and the final well defined CMC surface that creates the universal horizon in Schwarzchild and Kruskal-Szekeres coordinates. Once a conformal compactification has been performed the casual structure is clear in Figure \ref{fig:plot0pn} with the additional sub-UH CMC surfaces (which end in singularity, rather than the space-like infinity $i^0$).

\subsubsection{Case 2: $Q\neq0$ \& $b\leq0$}
\begin{figure}[t]
  \centering
    \includegraphics[width=0.5\textwidth]{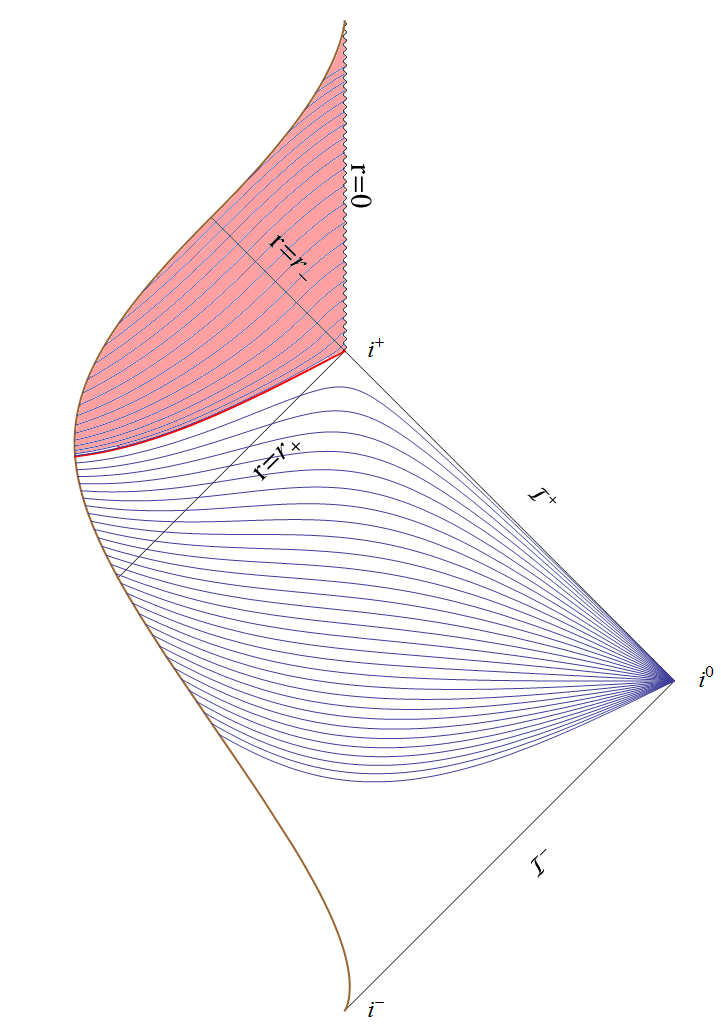}
  \caption{The Penrose diagram for $Q=0.99 M$ and $\epsilon=1$ collapsing shell depicting the UH. Coloured lines/region have the same meaning as Figure \ref{fig:plot0log}}   with the inclusion of sub-UH CMC surfaces in blue.
  \label{fig:plotqpn}
\end{figure}

For $\epsilon$ small enough such that the numerator of $b$ remains negative, the shell is unable to rebound before collapsing to a singularity. In Schwarzchild and the charged generalization of Kruskal-Szekeres remain almost identical in their analogous charts. The causal structure in Figure \ref{fig:plotqpn} reveals the distinction from case 1. The collapse ends in the coordinates, colloquially called the {\it parallel universe}. 

\subsubsection{Case 3: $Q\neq 0$ \& $0<b<b/\epsilon+1/\epsilon^2$}
\begin{figure}
  \centering
    \includegraphics[width=0.5\textwidth]{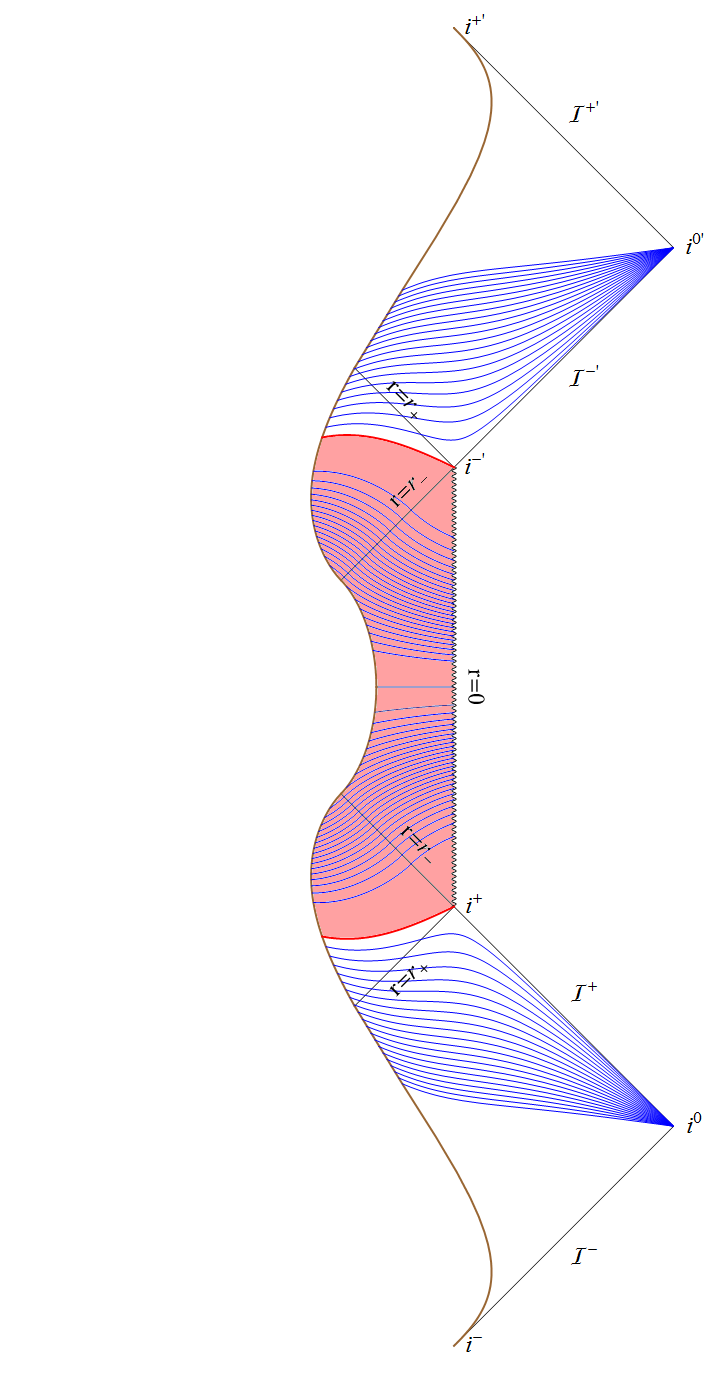}
  \caption{The Penrose diagram for  $Q=0.99 M$ and $\epsilon=1.1$ values. These parameters make $b/(\epsilon-1)<b/\epsilon+1/\epsilon^2$. Coloured lines/region have the same meaning as Figure 
\ref{fig:plot0log}  with the inclusion of sub-UH CMC surfaces in blue.
  } 
  \label{fig:UHRN3}
\end{figure}

For the range of $\epsilon$ such that $0<b<\frac{M}{\epsilon^2}$, the shell rebounds at the radius of $\frac{b}{\epsilon-1}$ but in the parallel coordinates which distinguishes it from the next case. In particular this means that $t(R)$ has a stationary point between $r_+$ and $r_-$. Figure \ref{fig:UHRN3} shows the collapse and the corresponding causal structure for this case.

\begin{figure}
  \centering
    \includegraphics[width=0.5\textwidth]{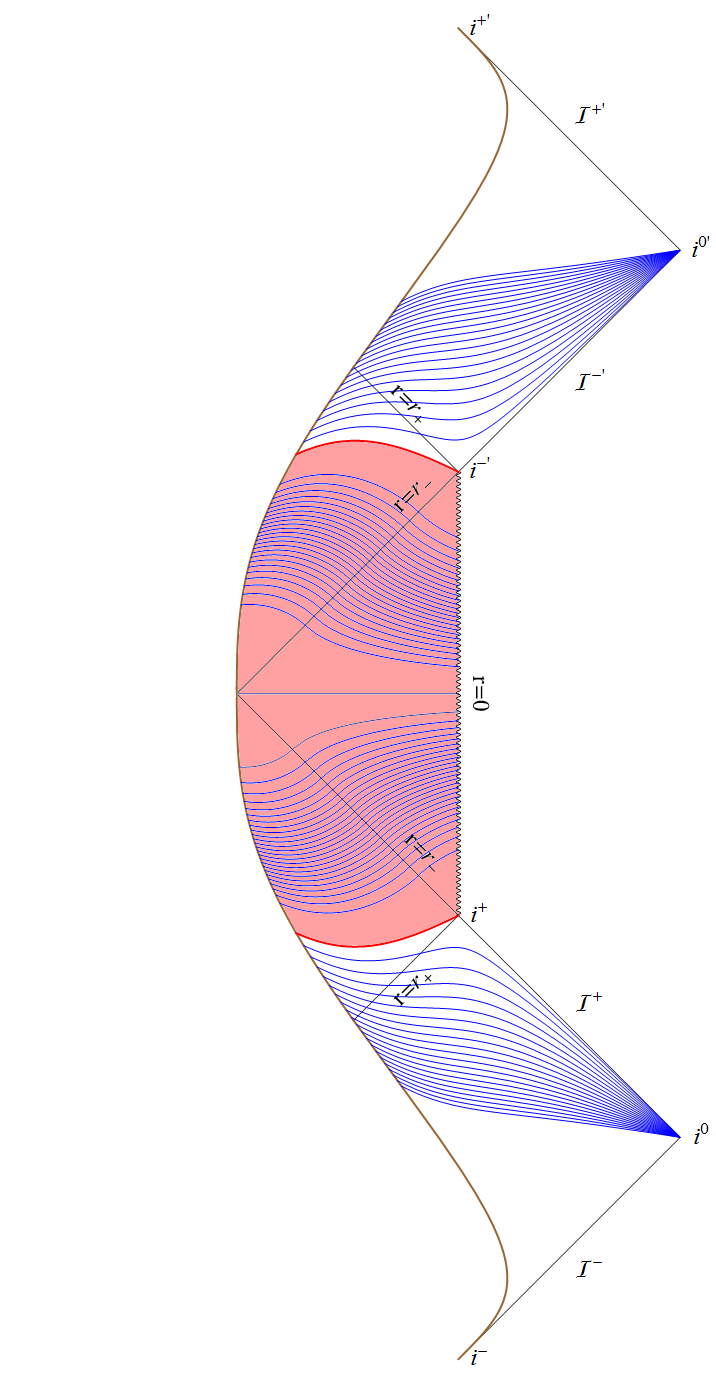}
  \caption{The Penrose diagram for  $Q=0.99M$ and $\epsilon$ set to make $b/(\epsilon-1)=b/\epsilon+1/\epsilon^2$. Coloured lines/region have the same meaning as Figure 
\ref{fig:plot0log}  with the inclusion of sub-UH CMC surfaces in blue.
  } 
  \label{fig:UHRN4}
\end{figure}

\begin{figure}
  \centering
    \includegraphics[width=0.5\textwidth]{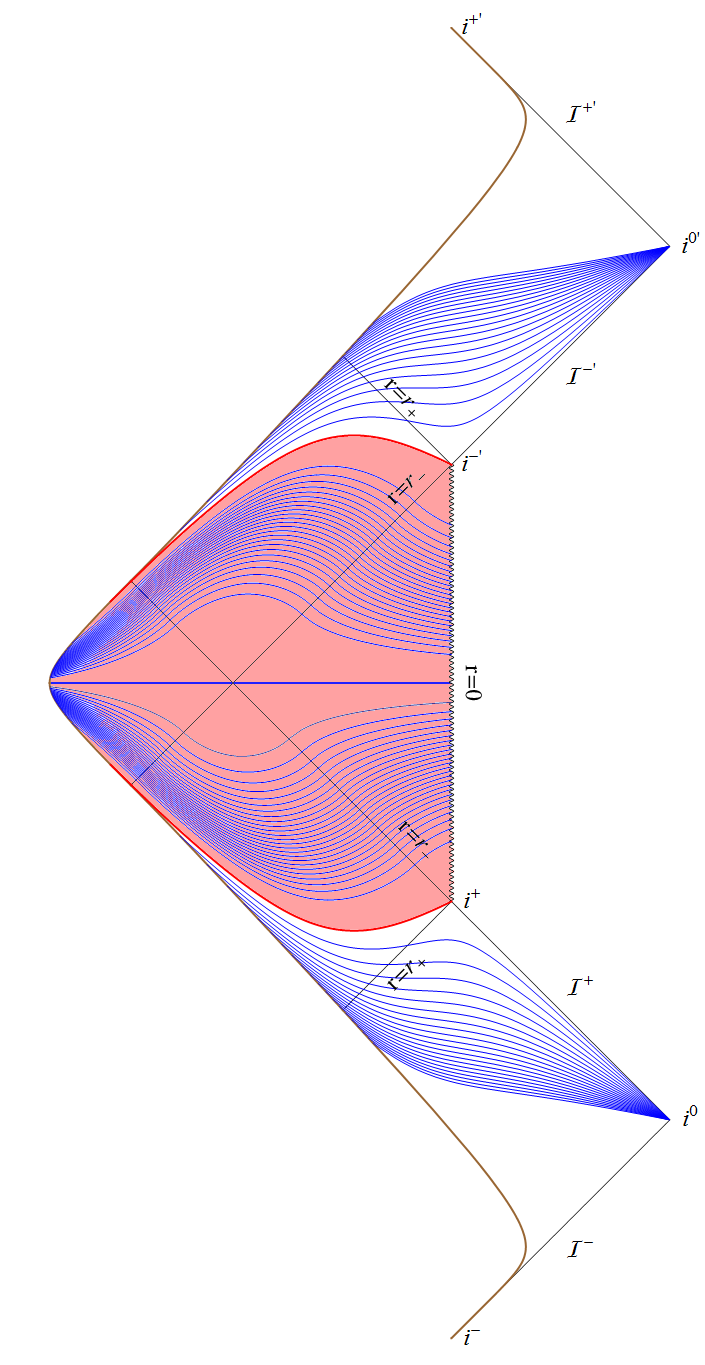}
  \caption{The Penrose diagram for  $Q=0.99 M$ and $\epsilon=3/2$ values. These parameters make $b/(\epsilon-1)>b/\epsilon+1/\epsilon^2$. Coloured lines/region have the same meaning as Figure 
\ref{fig:plot0log}  with the inclusion of sub-UH CMC surfaces in blue.
  } 
  \label{fig:UHRN2}
\end{figure}

\subsubsection{Case 4: $Q\neq0$ \& $b\geq b/\epsilon+1/\epsilon^2$}
Subsequently for $b>\frac{M}{ep}$ shell rebounds at the radius of $\frac{b}{\epsilon-1}$ in the original coordinate charts or exactly where the original and parallel coordinates meet . In particular this means that $t(R)$ has a stationary point at or inside $r_-$. The Schwarzchild and Kruskal-Szekeres coordinates are again nearly indistinguishable from case 1  except when the placement of $R_{lc}$ requiring that the UH being in a second charts however this does not reveal any new structure. Figure \ref{fig:UHRN4} and \ref{fig:UHRN2}  represents paths within this case,  $b=b/\epsilon+1/\epsilon^2$ and $b>b/\epsilon+1/\epsilon^2$ respectively. In this final case $R_{lc}<r_-$ resulting in the the UH piercing the $r_-$. Nevertheless, the singularity and the parallel interior horizon, which is considered to be unstable \cite{Poisson:1989zz} is still hidden within the UH.

\subsection{Censorship in Reissner-Nordstrom}	

Ultimately, in all the above cases, the UH shields any singularity from being probed even using superluminal signals and preserves a sense of cosmic censorship in Lorentz violating theories. It is apparent from the structure that every CMC is terminated at $i^0$, $i^+$, ${i^{-}}'$ ,${i^{0}}'$ or the singularity. We would posit an analogous, although informally made, statement to the original {\it weak} cosmic censorship conjecture: the set of points which can be connected to $i^0$ with CMC surfaces (an analogous property of being in the causal past of $\mathcal{I}^+$) is distinct from the set of points which can be connected to the singularity. Moreover, the boundary between these two sets will exactly be the universal horizon. 

Even though we have plotted the  maximal foliation of spacetime beyond the universal horizon, one can argue that the self-consistent evolution of the Lorentz-violating theory (including, e.g., backreaction or quantum effects, which we have ignored) stops at the universal horizon, which can be viewed as the boundary of classical spacetime (see Sec. VI in \cite{Saravani:2013} for more discussions).  The fact that both curvature singularity and the (potentially unstable) inner killing horizon \cite{Poisson:1989zz} lie beyond this region, further suggests a notion of {\it strong} cosmic censorship. 

Note that in the cases where a parallel universe exists, a second UH acts as a white hole horizon for superluminal signals.





\section{Spinning Black Holes: A Tale of Three Horizons}\label{spinningBH}
\subsection{Geometrical Definition of Universal Horizon}\label{geoUH}

In the previous section, we discussed the formation of universal horizon in a dynamic RN geometry. Before moving on to the spinning black hole case, it would be illuminating to acquire more intuition about the geometric nature of the universal horizon. We start by asking the following question: is there a way of finding the universal horizon in the final geometry (after collapse completed) without knowing the details of collapse?

 Let's consider the Schwarzschild case ($Q=0$ collapse). CMC surfaces in the thin shell collapse geometry describe the surfaces of constant global time. As we discussed earlier, as long as we are interested in the behaviour of these surfaces near the black hole (small radii) we can treat them as maximal surfaces ($K=0$). Maximal surfaces in this geometry inside the Schwarzschild radius asymptote to $r=\frac{3}{2}M$ before escaping to infinity. This suggests that $r=\frac{3}{2}M$ itself should be a maximal surface. In fact, one can simply verify that $r=r_*$ is a maximal (space-like) surface in Schwarzschild spacetime, only if $r_*=\frac{3}{2}M$. 

This observation suggests a geometrical definition for (asymptotic) universal horizon in stationary spacetimes; it is a maximal space-like hypersurface which is invariant under the flow of time-like killing vector. Let's discuss each element of this definition. 

First of all, UH has to be a maximal surface as we described earlier. It also has to be space-like, since it describes a constant global time surface. Secondly, it is invariant under time translation, as it is the asymptotic surface of the maximal slicing. 

We will show later explicitly that this definition does not pick a unique hypersuface. However, this should not be surprising, since the position of universal horizon depends on the behaviour of the background incompressible fluid (which defines the global time), unlike the killing horizon where its position is independent of the behaviour of the background fields \cite{Saravani:2013,Babichev:2007dw}.
Again, let's consider Schwarzschild spacetime. If we use the given definition of UH {\it with the additional assumption of spherical symmetry}, there is a unique solution of $r=\frac{3}{2}M$. However, there are many non-spherical UHs in the same geometry. 

Before moving to the spinning case, let's find the spherical UH of RN geometry using the definition given above. Assume $r=r_h$ to be the universal horizon and $v_{\mu}$ the unit normal vector to this surface. Solving 
\beq
\nabla_{\mu} u^{\mu}=0,
\eeq
we get
\beq\label{RNUH}
f'(r_h)=-4f(r_h),
\eeq
 with $f(r)=1-\frac{2M}{r}+\frac{Q^2}{r^2}$. Eq. (\ref{RNUH}) has a unique solution, which coincides with our previous result for universal horizon (\ref{r_h_ns}), and is plotted in Figure (\ref{fig:uni-hor}).
One can also directly check that \eqref{RNUH} is equivalent to system of equations \eqref{eq1_rh} and \eqref{eq2_rh}.

\subsection{Universal Horizon in Kerr geometry}\label{KerrUH}

In this section, we find the asymptotic universal horizon of Kerr metric. Given our definition above, it is a static axisymmetric\footnote{we assume that the background incompressible field obeys the axial symmetry of Kerr geometry.} (space-like) maximal surface. We express the Kerr metric in the following coordinates:
\bea
ds^2=&&\left(1-\frac{2mr}{\rho^2}\right)dt^2-\frac{\rho^2}{\Delta}dr^2-\rho^2d\theta^2-\frac{4mra\sin^2\theta}{\rho^2}dt d\phi\notag\\
&&-\left(r^2+a^2+\frac{2mra^2 \sin^2\theta}{\rho^2}\right)\sin^2\theta d\phi^2
\eea
where 
\bea
\rho^2&=&r^2+a^2\cos^2\theta,\\
\Delta&=&r^2-2mr+a^2.
\eea
The inner ($r_-$) and outer ($r_+$) killing horizons are the solution to $\Delta=0$. 

UH is the surface
\beq
r=r_h(\theta)
\eeq
which satisfies
\beq\label{maximal_Kerr}
\nabla_\mu v^\mu=0
\eeq
where $v^\mu$ is the (time-like) normal vector to the universal horizon. In other words,
\beq
v_\mu=\frac{1}{N}(0,1,-r_h',0)
\eeq
where $'$ is the derivative w.r.t $\theta$ and $N$ is the normalization factor 
\beq\label{Normalization}
N^2=-\frac{1}{\rho^2}\left(\Delta+r_h'^2\right).
\eeq
Equation \eqref{Normalization} leads to the following conclusion: demanding UH to be a space-like surface ($v^\mu$ to be time-like) requires the UH to be positioned between the inner and the outer killing horizons
\beq
N^2>0\to \Delta<0\to r_-<r_h(\theta)<r_+.
\eeq

\begin{figure*}[t]
\begin{subfigure}[t]{0.49\textwidth}
\centering
\includegraphics[width=\hsize]{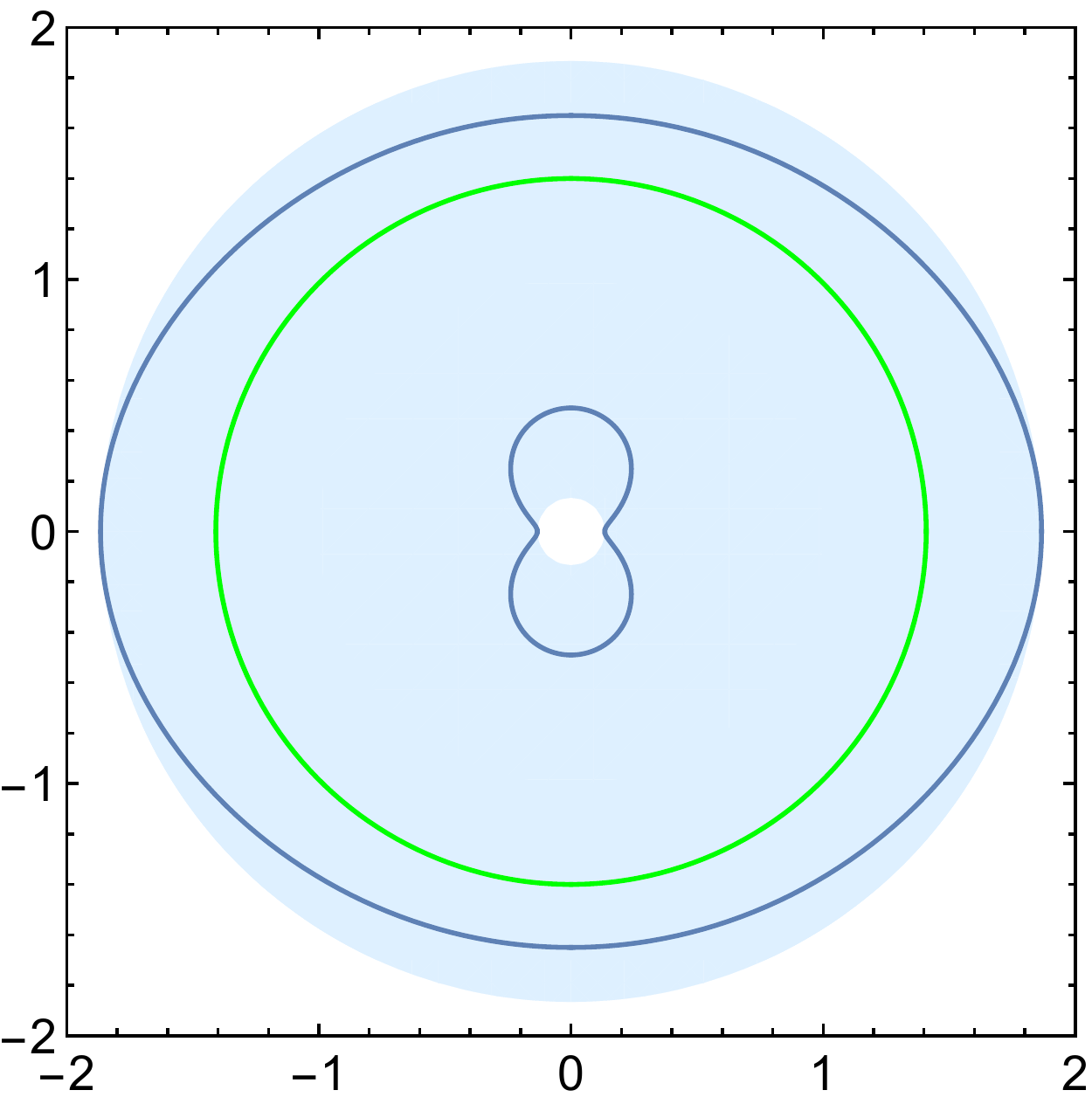}
\caption{$a=\frac{m}{2}$}
\end{subfigure}
\hfill
\begin{subfigure}[t]{0.49\textwidth}
\centering
\includegraphics[width=\hsize]{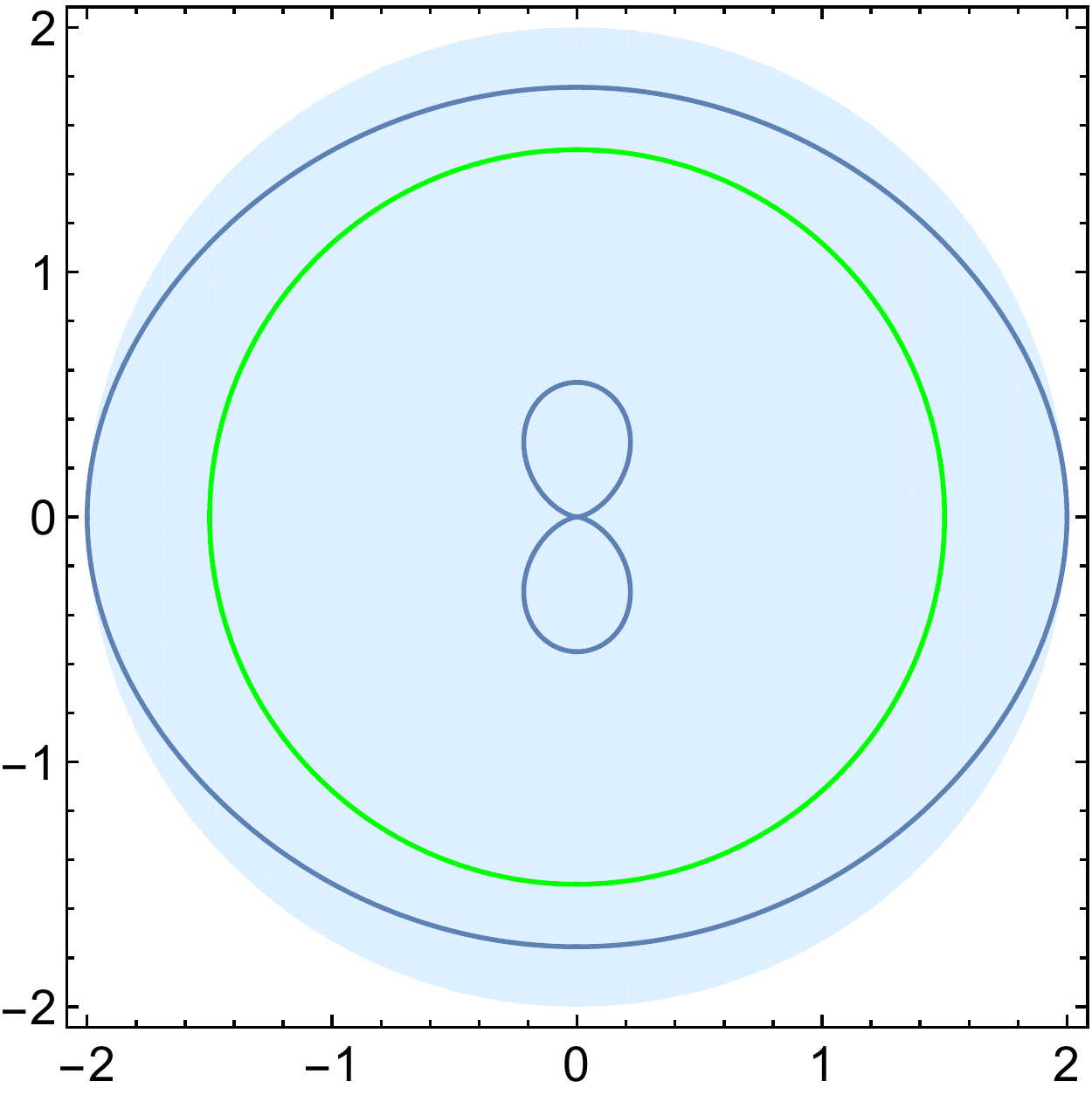}
\caption{$a=0$ (Schwarzschild)}
\end{subfigure}
\caption{Polar plot $\frac{r(\theta)}{m}$. Green curve is the UH solution of \eqref{perturb_sol}. Blue curves are additional numerical solutions to \eqref{master_eq} and shaded region is the region between inner and outer killing horizons. The outer (inner) UH is tangent to the outer (inner) killing horizon at $\theta=\frac{\pi}{2}$.}
\label{axisymmetric_UH}
\end{figure*}

Now on to finding $r_h(\theta)$: Equation \eqref{maximal_Kerr} takes the form 
\bea\label{master_eq}
&2(r_h-m)+\frac{r_h(r_h^2-2mr_h+a^2)}{r_h^2+a^2\cos^2\theta}-\frac{(r_h-m)(r_h^2-2mr_h+a^2)}{r_h^2-2mr_h+a^2+r_h'^2}
\notag\\
&=\frac{r_h'}{\tan\theta}+r_h''-r_h'\left[\frac{a^2\sin\theta\cos\theta}{r_h^2+a^2\cos^2\theta}+\frac{r_h'r_h''}{r_h^2-2mr_h+a^2+r_h'^2}\right].
\eea
One way to find the solution of this differential equation is to expand $r_h(\theta)$ in powers of $a$
\beq\label{perturb_sol}
r_h(\theta)=m\sum_{n=0}\frac{a^n}{m^n}r^{(n)}(\theta)
\eeq
and solve the differential equation order by order. At zero order ($a=0$), we expect $r^{(0)}=\frac{3}{2}$. At any higher order, we find Legendre differential equation. Requiring finite solution at $\theta=0$ and $\theta=\pi$, this gives us a unique solution at any order. Here is the solution up to the order $a^{4}$:
\bea
r^{(2n-1)}&=&0,~~ n\in\{1,2,\cdots\}\notag\\
r^{(0)}&=&\frac{3}{2}\notag\\
r^{(2)}&=&-\frac{1}{36} \cos ^2\theta-\frac{13}{36}\notag\\
r^{(4)}&=&\frac{49}{10692}\cos^4\theta+\frac{29 }{4752}\cos ^2\theta-\frac{1057}{14256}.\notag
\eea
Surprisingly though, upon solving \eqref{master_eq} numerically, we have found two more solutions that are different from \eqref{perturb_sol} and  do not approach to $r_h=\frac{3}{2}m$ as $a \rightarrow 0$ (see Figure \ref{axisymmetric_UH}). 

The case of $a=0$ is interesting, since the background geometry is spherically symmetric, and yet we have found two axisymmetric UHs. Moreover, we can always perform a rotation and get two other axisymmetric UHs. This means that there are two families (with infinite number in each family) of axisymmetric UHs in Schwarzschild spacetime.



\section{Conclusion}\label{conclude}

In the incompressible (or infinitely fast  propagation speed) limit of many Lorentz-violating theories of gravity, surfaces of constant mean curvature define the preferred foliation \cite{Saravani:2013}\footnote{For a careful study of theories with infinite sound speed see \cite{Bhattacharyya:2015gwa}.}. For such theories, one may worry that superluminal signal propagation may lead to naked singularities. In this paper, we have shown that a {\it universal} horizon always forms when a charged spherical shell collapses to form a Reissner--Nordstrom black hole. Evidence that causal horizon formation will take place in Lorentz-violating theories supports a conjecture similar to cosmic censorship in General Relativity.

We see that the universal horizon acts almost like an extension of $i^+$, since any observer approaching the UH will pass through all future CMC surfaces outside the UH. Consequently, the analysis conducted here is likely only valid in the classical regime (ignoring quantum effects like the evaporation of BH).  As a result the region close to the UH is likely where non-classical effects will begin to become relevant. Making claims past this region may require the full UV theory. 

We have also presented a geometric definition for the UH which provides a tool for finding generic solutions in non-spherically symmetric geometries.  This tool is additionally valuable as the full evolution of the system up to the point of UH formation is not needed to be explored. In particular, we show how the definition can be applied to the Kerr geometry, revealing a family of solutions in a non-spherical geometry. 

Additional horizon solutions may be real solutions and the result of different (possibly more generic) collapse histories; or just an artifact of our definition which does not single out the correct solution. This further motivates numerical dynamical studies of {\it non-spherical} collapse in Lorentz-violating gravitational theories, as the spherical solution (and universal horizon \cite{Blas:2011ni}) may be unstable. 

Gravitational dynamics within real black holes may yet have more surprises in store for us!

\bibliographystyle{ieeetr}
\bibliography{Universal.Horizon_v2}

\appendix

\end{document}